\documentclass[prd,aps,twocolumn,showpacs,preprintnumbers,amsmath,amssymb,nofootinbib]{revtex4}
\usepackage[dvips]{graphicx}
\usepackage{subfig}
\usepackage{epsf}
\usepackage{amsmath}
\usepackage{amssymb}
\usepackage{graphicx}
\usepackage{dcolumn}
\usepackage{bm}
\begin{document}

\title{Breaking parameter degeneracy in interacting dark energy models from observations}
\author{Xiao-Dong Xu, } \affiliation{Department of Physics, Fudan University, 200433 Shanghai, China}
\author{Jian-Hua He, Bin Wang}   \affiliation{INPAC and Department of Physics, Shanghai Jiao Tong University, 200240 Shanghai, China}

\begin{abstract}
We  study the interacting dark energy model with time varying dark
energy equation of state. We examine the stability in the
perturbation formalism and the degeneracy among the coupling between
dark sectors, the time-dependent dark energy equation of state and
dark matter abundance in the cosmic microwave background radiation.
Further we discuss the possible ways to break such degeneracy by
doing global fitting using the latest observational data and we get
a tight constraint on the interaction between dark sectors.

\end{abstract}

\pacs{98.80.Cq} \maketitle

\section{Introduction}
It is convincing that cold dark matter (DM) and dark energy (DE) are
two dominant sources governing the evolution of our universe.
Considering that these two components are independent is a very
specific assumption, in the framework of field theory  it is more
natural to consider the inevitable interaction between them
\cite{secondref}. An appropriate  interaction in the dark sectors
can alleviate the coincidence problem and understand why energy
densities of dark sectors are of the same order of magnitude today
\cite{10}-\cite{14}.

However, it was suspected that the coupling between dark sectors might lead to the curvature perturbation instability \cite{x}. This problem was
later clarified in \cite{y} where it was shown that the stability of curvature perturbation holds by appropriately choosing the forms of the
interaction and the values of the constant equation of state (EOS) of the DE. Another possibility to cure the instability was suggested in
\cite{z}.

The observational signature of the interaction between dark sectors
has been widely discussed \cite{10}-\cite{AbdallaPLB09}. It was
found that the coupling in dark sectors can affect significantly the
expansion history of the universe and the growth history of
cosmological structures. A number of studies have been devoted to
grasp the signature of the dark sectors mutual interaction from the
probes of the cosmic expansion history by using the WMAP, SNIa, BAO
and SDSS data etc \cite{71}-\cite{pp}. Interestingly it was
disclosed that the late ISW effect has the unique ability to provide
insight into the coupling between dark sectors \cite{hePRD09}.
Furthermore, complementary probes of the coupling within dark
sectors have been carried out in the study of the growth of cosmic
structure \cite{31}-\cite{AbdallaPLB09}. It was found that a
non-zero interaction between dark sectors leaves a clear change in
the growth index \cite{31,Caldera,peacock}. In addition, it was
argued that the dynamical equilibrium of collapsed structures such
as clusters would acquire a modification due to the coupling between
DE and DM \cite{pt,AbdallaPLB09}, which could leave the signature in
the virial masses of clusters \cite{AbdallaPLB09}. The imprint of
dark sectors interaction in the cluster number counts was disclosed
in \cite{bb}. N-body simulations of structure formation in the
context of interacting DE models were studied in \cite{Baldi}.

Since both DE and DM are currently only detected via their gravitational effects and any change in the DE density is conventionally attributed to
its equation of state $w$,  there is an inevitable degeneracy while extracting the signature of the interaction between dark sectors and other
cosmological parameters. The degeneracies among the dark sector coupling, the the equation of state (EoS) of DE and the DM abundance were first
discussed in \cite{he2010}. In the formalism of the perturbation theory, it was found that the degeneracy can be broken and tighter constraint on
the interaction between dark sectors can be obtained from observations.

In \cite{he2010}, the EoS of DE was taken to be constant. Recently,
more accurate data analysis tells us that the time varying DE gives
a better fit than a cosmological constant \cite{Yin9}. The time
varying EoS influences a lot on the universe evolution, perturbation
stability and it must affect the constraint on the interaction
between dark sectors.  Thus it is of great interest to generalize
the discussion in \cite{he2010} to the interacting DE model with
time-dependent DE EoS. The observational constraints on an
interacting DE with time varying EoS were discussed in
\cite{75,beans}. In this work we will first discuss the stability of
the perturbation and the degeneracy between the dark sector
interaction and the time varying EoS of DE. Furthermore we will
explore possible ways to break the degeneracy among the dark
sectors' coupling, the time-dependent DE EoS and the DM abundance.
In our study we will use the popular Chevallier-Polarski-Linder
(CPL) parametrization \cite{Gong9} to describe the time varying DE
EoS, which expresses the EoS in terms of the scale factor in the
form $w(a) = aw_0 + (1-a)w_e$. In the early time, $a \ll 1$, $w
\simeq w_e$. At the present time when $a\simeq 1$, $w\simeq w_0$.

\section{perturbation formalism and its stability}

In the spatially flat Friedmann-Robertson- Walker(FRW) background, if there is an interaction between DE and DM, neither of them can evolve
independently. The (non)conservation equations are described by
\begin{eqnarray}
    \rho_{c}' + 3 \mathcal{H} \rho_c = a Q_c, \\
    \rho_{d}' + 3 \mathcal{H} (1 + w) \rho_d = a Q_d,
\end{eqnarray}
where the subscript c represents DM and d stands for DE. $Q_{\lambda}$ is the term leading to energy transfer. Considering that there is only
energy transfer between DE and DM, we have $Q_c=-Q_d=Q$. The sign of $Q$ determines the direction of the energy transfer. For positive $Q$, the
energy flows from DE to DM. For negative $Q$, the energy flow is reversed. In our following study, we adopt the phenomenological interaction in
the form $Q = 3 \xi H \rho_c$, where $\xi$ is the coupling constant. This merely repeats the analysis of [10], where the energy transfer was
assumed to be proportional to the energy density, but here we have the energy density times the Hubble parameter. In [7], it was indicated that
qualitatively similar conclusions apply to both cases. The presence of DM energy density in the interaction was shown problematic in the
stability of perturbation \cite{x}. It was argued that only for constant EoS of DE in the phantom region, this interaction form can be viable and
effective to alleviate the coincidence problem\cite{y}. It is of interest to explore this interaction form when the DE EoS is time-dependent with
the CPL parametrization, especially when the DE EoS is in the quintessence region.

As discussed in \cite{he2010}, the coupling vector is specified in co-moving frame as
\begin{equation*}
Q^{\nu}_{(\lambda)} = (\frac{Q_{(\lambda)}}{a}, 0, 0, 0)^T.
\end{equation*}
The perturbed form of the zero component $\delta Q^{0}_{(\lambda)}$ can be uniquely determined from the background energy-momentum transfer and has
the form
\begin{equation*}
\delta Q^{0}_{(\lambda)} = -\Psi Q_{(\lambda)}/a + \delta Q_{(\lambda)}/a.
\end{equation*}
While the spatial component of the perturbed energy-momentum transfer $\delta Q^{i}_{(\lambda)}$ can be set to zero since there is no
non-gravitational interaction in the DE and DM coupled system and only the inertial drag effect appears in the system due to the stationary
energy transfer between DE and DM as discussed in \cite{he2010}.

The perturbation equations for DM and DE have the forms \cite{he2010, 31}
\begin{align*}
D_c' &= -kU_c + 3\mathcal{H}\xi\Psi - 3\xi\Phi'\quad, \\
U_c' &= -\mathcal{H}U_c + k\Psi - 3\mathcal{H}\xi U_c \quad;\\
D_d' &= -3\mathcal{H}(C^2_e - w)D_d - 9\mathcal{H}^2(C^2_e - C^2_a)\frac{U_d}{k} \\
     &+ \{3w' - 9\mathcal{H}(w - C^2_e)(\xi r + 1 + w)\}\Phi \\
     &+ 3\xi r \Phi' - 3\mathcal{H}\xi r\Psi + 3\mathcal{H}\xi r(D_d - D_c) \\
     &- 9\mathcal{H}^2(C^2_e - C^2_a)\xi r \frac{U_d}{(1+w)k} - kU_d \quad, \\
U_d' &= -\mathcal{H}(1-3w)U_d - 3C^2_e(\xi r + 1 + w)k \Phi \\
     &+ 3\mathcal{H}(C^2_e - C^2_a)\xi r \frac{U_d}{1+w} + 3\mathcal{H}(C^2_e - C^2_a)U_d \\
     &+ kC^2_e D_d + (1+w)k\Psi + 3\mathcal{H}\xi r U_d.
\end{align*}
$r = \rho_c / \rho_d$ is the ratio of DM to DE. In the above, $C^2_a = w < 0$. However, it is not clear what expression we should  have for
$C^2_e$ . In \cite{x} it has been argued in favor of $C^2_e = 1$. This is correct for the scalar field, but it is not obvious for other cases.
The most dangerous possibility, as far as instabilities are concerned, is $C^2_e = 1 \neq C^2_a = w < 0$ since the first term in the second line
of the equation of $U_d$ can lead to the blow up when $w$ is close to $-1$. Assuming that $C^2_e = 1, C^2_a = w$, the last two equations above
for DE can be rewritten as
\begin{align*}
D_d' &= 3\mathcal{H}(-1+ w+\xi r)D_d - 9\mathcal{H}^2(1-w) (1 + \frac{\xi r}{1+w}) \frac{U_d}{k} \\
     & -kU_d + \{3w' + 9\mathcal{H}(1-w)(\xi r +1+w)\} \Phi + 3\xi r\Phi' \\
     & -3\mathcal{H}\xi r\Psi - 3\mathcal{H}\xi r D_m \quad,\\
U_d' &= -2\mathcal{H}(1+\frac{3\xi r}{1+w})U_d +kD_d- 3(\xi r+1+w)k\Phi \\
     &+(1+w)k\Psi.
\end{align*}
Numerically, we find that the first two terms on the RHS of the
perturbation equations for DE contribute more than other terms to
the divergence. In order to explain the reason for the blow-up, we
keep the leading terms and use $\xi r\sim -w$ in the early universe
\cite{y}. In the early universe $w\sim w_e$ and the above equations
can be reduced in the form
\begin{align*}
D_d' &= -3\mathcal{H}D_d - 9\mathcal{H}^2 \frac{1-w_e}{1+w_e}\frac{U_d}{k}\quad,\\
U_d' &= 2\mathcal{H}\frac{1-2w_e}{1+w_e}U_d +kD_d.
\end{align*}
The second order differential equation for $D_d$ can be written as
\begin{equation*}
D''_d = \bigg( 2\frac{\mathcal{H}'}{\mathcal{H}} -\frac{1+7w_e}{1+w_e}\mathcal{H} \bigg) D'_d+3(\mathcal{H}'-\mathcal{H}^2)D_d.
\end{equation*}
In the radiation dominated period, we have $\mathcal{H}\sim 1/\tau, \mathcal{H}'\sim -1/\tau^2$, thus
\begin{equation*}
D''_d = -3\frac{1+3w_e}{1+w_e}\frac{D'_d}{\tau}-\frac{6}{\tau^2}D_d,
\end{equation*}
which gives the solution
\begin{equation*}
D_d = C_1\tau^{r_+}+C_2\tau^{r_-},
\end{equation*}
where $$r_{\pm}=-\frac{1+4w_e+\pm\sqrt{-5-4w_e+10w_e^2}}{1+w_e}.$$

\begin{figure}[hptb]
    \includegraphics[scale=0.8]{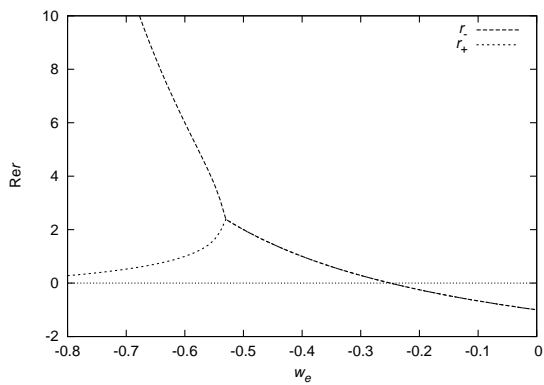}
    \caption{The Behavior of the indices $r_{\pm}$. $r_+$ and $r_-$ meet at the point
where the square root becomes zero. The real part of $r_{\pm}$ are the same on the right side of this point.}
    \label{w_index}
\end{figure}

When $r_\pm$ is a real positive value, the perturbation will blow
up. In \cite{y} for the constant DE EoS, $w$ was required to be
smaller than $-1$ to accommodate the stability. For the DE EoS with
CPL parametrization, in the early time $w\simeq w_e<-1$ can also
allow negative $r_\pm$ to provide stability. Moreover as shown in
Fig.\ref{w_index} when $w\simeq w_e>-1/4$ in the early time, the
stability can be safely protected as well.

\begin{figure}[hptb]
    \subfloat[]{
        \includegraphics[scale=0.8]{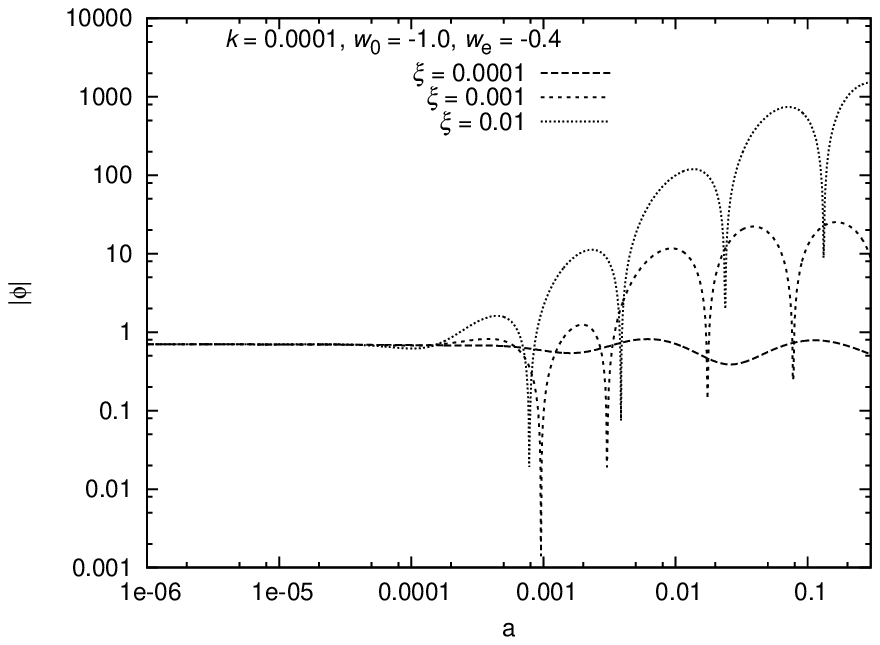}
    }\\
    \subfloat[]{
        \includegraphics[scale=0.8]{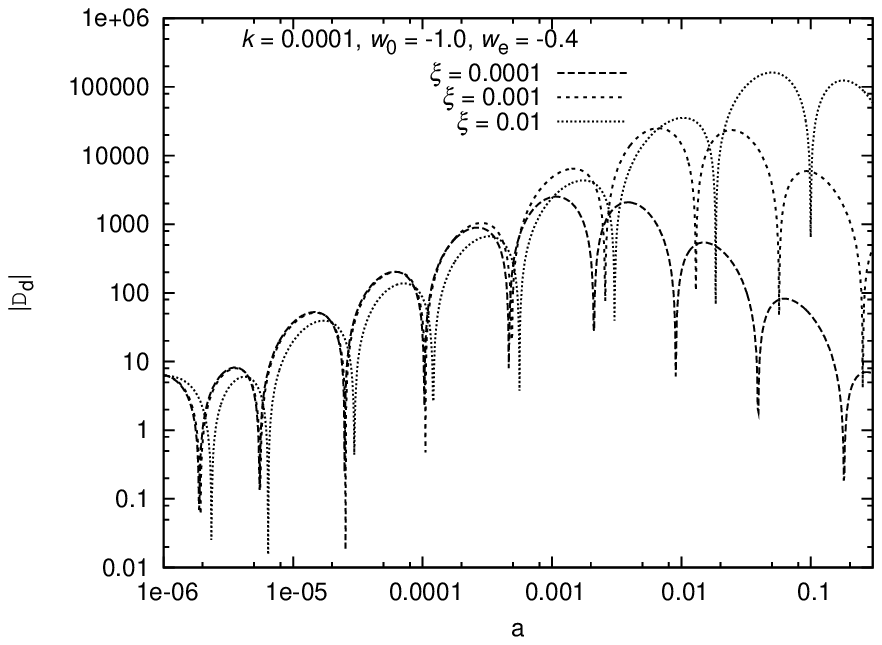}
    }
    \caption{The evolution of the curvature perturbation and DE density perturbation for
various  $\xi$ when $-1<w_e<-1/4$.}
    \label{pert}
\end{figure}

When $-1<w_e<-1/4$, the above approximate analytical analysis cannot
help, we need to count on the numerical calculation to see the
stability of perturbation in the early universe. Results are shown
in Fig.\ref{pert}. When $w_e$ falls in this range, the perturbation
grows slower for weaker coupling. When the existing dark sector
coupling is weak enough, it has the possibility to keep the
perturbation stable. For smaller $w_e$ in the range $-1<w_e<-1/4$,
weaker interaction is required to keep the stability in
perturbation.

Thus unlike the constant DE with $w>-1$, the time varying DE can allow stable perturbation when it is interacting with DM. However in some ranges
of $w$ in the quintessence region,  in order to keep the stability in the perturbation we have the upper limit on the strength of the
interaction. When the DE EoS is in the phantom region, the stability of the perturbation is always protected in the presence of the interaction
between dark sectors, which is the same as the case we observed for the constant EoS of DE \cite{y}.

\section{degeneracies of interaction, DE EoS and DM abundance in CMB spectrum}

\begin{figure}[hptb]
    \subfloat[$w_e \sim \xi$]{
        \includegraphics[scale=0.8]{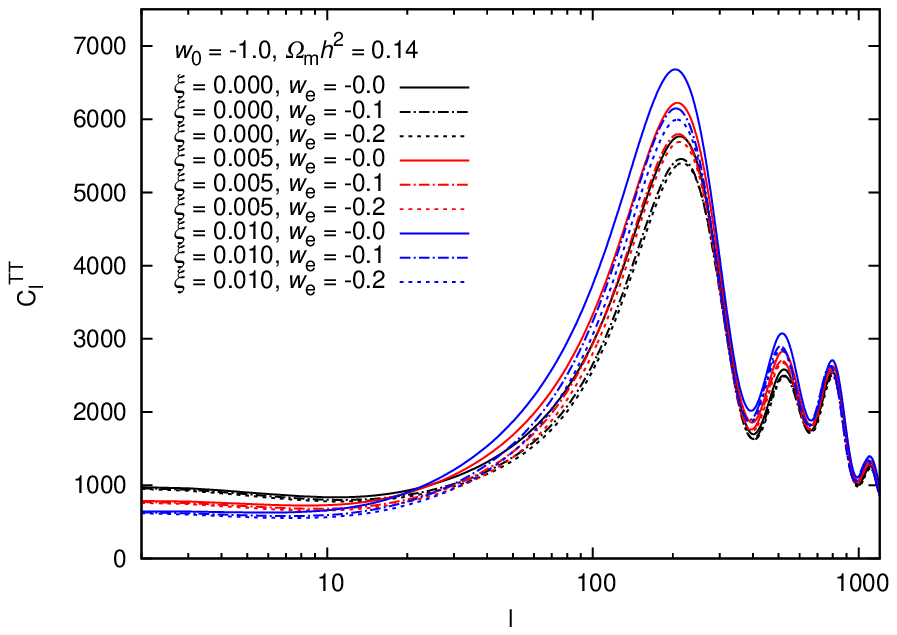}
    }
    \\
    \subfloat[$w_0 \sim \xi$]{
        \includegraphics[scale=0.8]{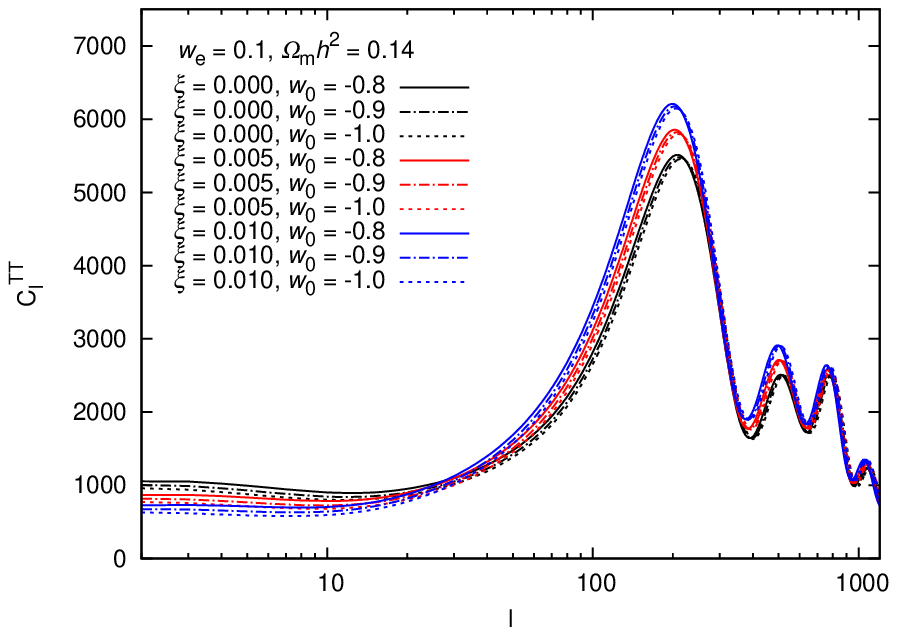}
    }
    \caption{The dependences of CMB angular power spectrum on the time varying DE EoS and the coupling between dark sectors.}
    \label{deg}
\end{figure}

With the perturbation formalism at hand, we can study the  influence of the interaction between dark sectors and other cosmological parameters on
the CMB power spectrum. We will concentrate our attention on the DE EoS in the quintessence region $w>-1$, since it has different property as
discussed in the above section compared with the case with constant EoS. In Fig.\ref{deg} we illustrate the theoretical computation results of the CMB
power spectrum for changing $w_0, w_e$ in the CPL parametrization of DE EoS and coupling constant $\xi$ but with DM abundance fixed.

We see in the CMB TT angular power spectrum that the change of the
$w_e$ only modifies the acoustic peaks, while it does not influence
the low-l part of the spectrum. The influence of $w_0$ is on the
contrary, it changes little of the acoustic peaks, but influences
more on the small-l spectrum. When $w_0$ decreases, low-l spectrum
will be suppressed. Different effects given by $w_e$ and $w_0$ on
CMB spectrum are interesting. Since we have more CMB data in large-l
than small-l spectrum, it is easier to constrain $w_e$ than $w_0$.
This can be used to explain the fitting results in the following.

The above discussion is for the fixed coupling constant. If we allow the change of $\xi$, we see that at the acoustic peaks, $\xi$ plays more
important role in the change of the spectrum than the influence of $w_e$. In the low-l region, we also observe that $\xi$ effect is more
important than $w_0$ to suppress the CMB spectrum.

\begin{figure}[hptb]
    \includegraphics[scale=0.8]{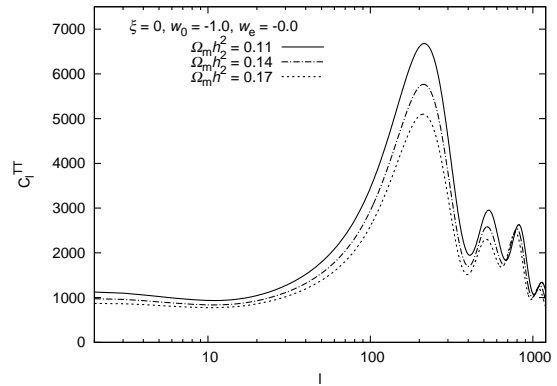}
    \caption{The dependence of CMB angular power spectrum on the DM abundance.}
    \label{deg2}
\end{figure}

The above discussion is valid for fixed DM abundance. Now we
investigate the dependence of CMB angular power spectrum on the
abundance of matter, $\omega_m=\Omega_m h^2$. Since the abundance of
baryon is fixed, it is identical to investigate the effect of the
abundance of CDM. Although the abundance of the DM does not affect
much on the low-l CMB power spectrum, it quite influences the
amplitude of the first and second acoustic peaks in CMB TT angular
power spectrum (see Fig.\ref{deg2}). Decreasing $\omega_m$ will
enhance the acoustic peaks. This effect is degenerated with the
influence given by the dark sectors' interaction and $w_e$ as we
observed in Fig.\ref{deg}. A possible way to break this degeneracy
is to consider the influence of the interaction on the low-l CMB
spectrum. Moreover, we can include further observations to get a
complementary constraint on the matter abundance and this in turn
can help to constrain the coupling between dark sectors.

In order to extract the signature of the interaction and put constraints on other cosmological parameters, we need to use the latest CMB data
together with other observational data. We report the results of fitting in the next section.

\section{fitting results}
In this section we confront our models with observational data by
implementing joint likelihood analysis. We take the parameter space
as
$$P=(h,\omega_b,\omega_{m},\tau,\ln[10^{10}A_s],n_s,\xi,w_0, w_e)$$
where $h$ is the hubble constant, $\omega_b=\Omega_bh^2, \omega_{m}=\Omega_{m}h^2$, $A_s$ is the amplitude of the primordial curvature
perturbation, $n_s$ is the scalar spectral index, $\xi$ is the coupling constant when we choose the interaction in proportion to the energy
density of DM. To avoid the negative energy density of DE in the early time of the universe\cite{heJCAP08}, we limit the coupling constant $\xi$
to be positive which indicates the energy flow from DE to DM. $w_0, w_e$ are parameters in the CPL parametrization of DE EoS. Since we
concentrate on the DE EoS in the quintessence region, we set the limit $-1<w_0<0, -1<w_e<0$ in our data fitting. We choose the flat universe with
$\Omega_k=0$ and our work is based on CMBEASY code\cite{easy}. The fitting results are listed in Table.\ref{bestfit}.

\begin{figure*}[hptb]
    \centering
    \subfloat[$h$]{
        \includegraphics[scale=0.8]{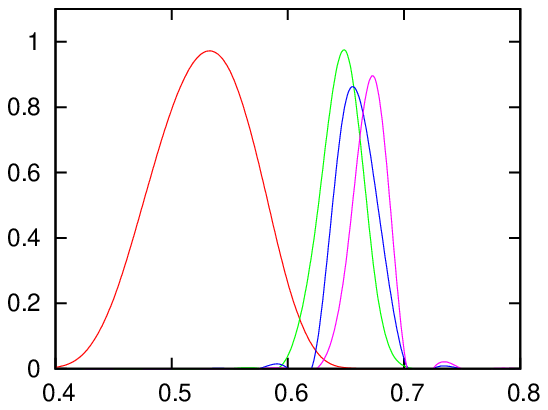}
    }
    \subfloat[$\Omega_m h^2$]{
        \includegraphics[scale=0.8]{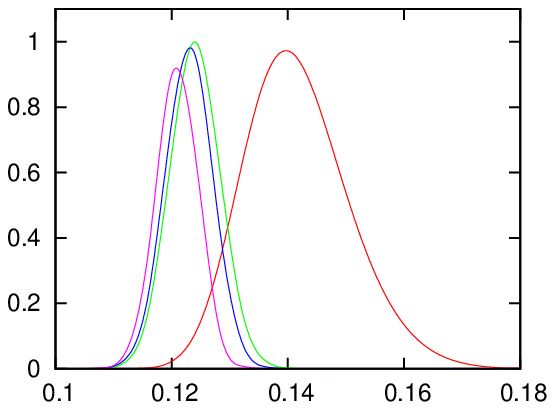}
    }
    \subfloat[$\Omega_b h^2$]{
        \includegraphics[scale=0.8]{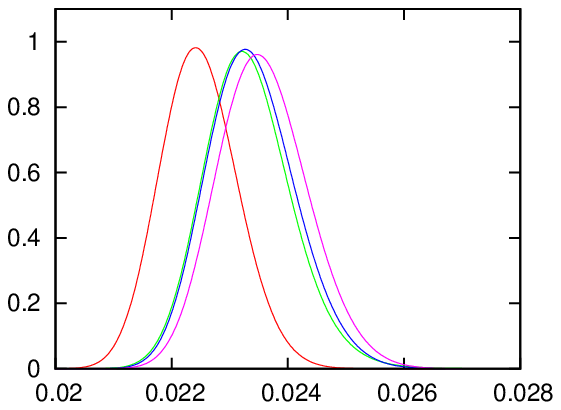}
    }\\
    \subfloat[$\tau$]{
        \includegraphics[scale=0.8]{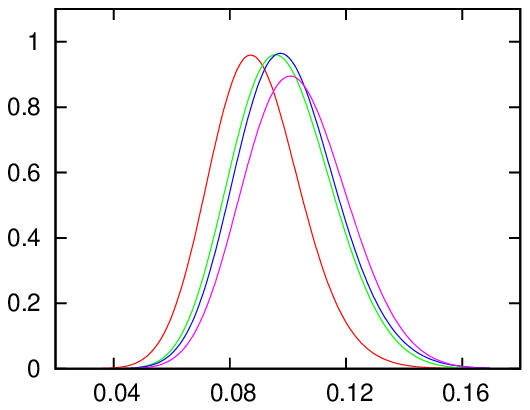}
    }
    \subfloat[$n$]{
        \includegraphics[scale=0.8]{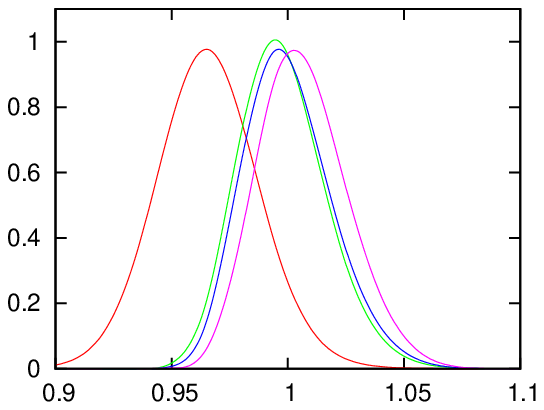}
    }
    \subfloat[${\rm ln}10^{10}A_s$]{
        \includegraphics[scale=0.8]{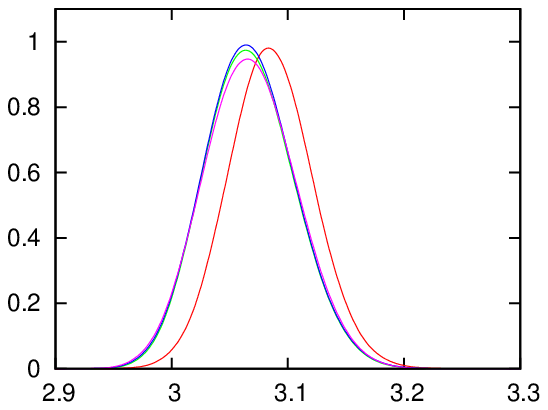}
    }\\
    \subfloat[$w_0$]{
        \includegraphics[scale=0.8]{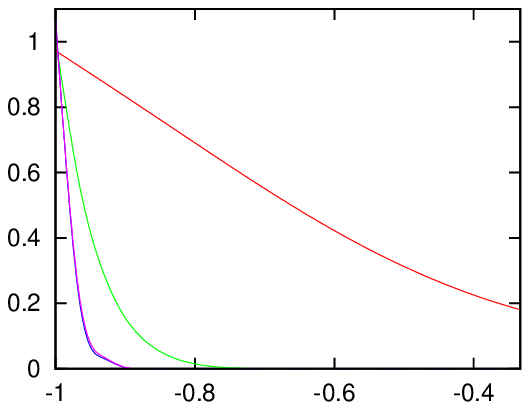}
    }
    \subfloat[$w_e$]{
        \includegraphics[scale=0.8]{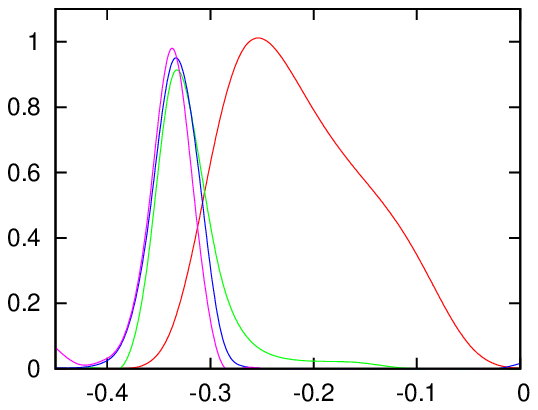}
    }
    \subfloat[$\xi$]{
        \includegraphics[scale=0.8]{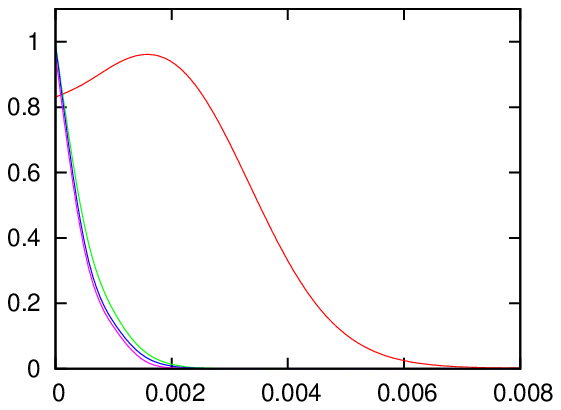}
    }\\
    \subfloat[]{
        \includegraphics[scale=1]{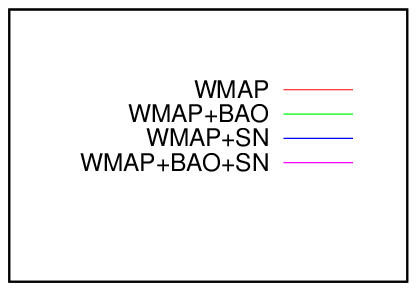}
    }
    \caption{1-D marginalized likelihoods for cosmological parameters. WMAP alone (red), WMAP+BAO(green), WMAP+SN(blue) and WMAP+BAO+SN(purple)}
    \label{1Dlike}
\end{figure*}

When we only use the CMB anisotropy data from the seven-year Wilkinson Microwave Anisotropy Probe (WMAP), in Fig.\ref{1Dlike} we show the 1d
marginalized likelihoods for all of the primary MCMC parameters of our model. It is clear that, just from CMB data, the constraint on $w_0$ is
worse than that on $w_e$. This is consistent with the analysis we did in the theoretical study. The $w_e$ affects more on the acoustic peaks.
More observational CMB data there can help to constrain $w_e$ well. For our model with the interaction between dark sectors in proportion to the
energy  density of DM, we find that CMB data alone can impose tight constraints on couplings and $\Omega_mh^2$. This can be understood by using
our theoretical analysis that the degeneracy between the coupling and the DM abundance can be broken by looking at the low-l CMB spectrum.

\begin{table*}[htbp]
\caption{The cosmological parameters from the global fitting.} \centering \label{bestfit}
\renewcommand\arraystretch{1.5}
\begin{tabular}{ccccc}
    \hline
    Parameter & WMAP & WMAP+BAO & WMAP+SN & WMAP+BAO+SN \\
    \hline
    $\Omega_m h^2$ & $0.140^{+0.009}_{-0.008}$ & $0.124^{+0.004}_{-0.004}$ & $ 0.123^{+0.004}_{-0.004}$ & $0.121^{+0.004}_{-0.004}$ \\
    $\Omega_b h^2$ & $0.0224^{+0.0007}_{-0.0006}$ & $0.0232^{+0.0008}_{-0.0007}$ & $ 0.0233^{+0.0008}_{-0.0007}$ & $0.0235^{+0.0008}_{-0.0007}$ \\
    $h$ & $0.532^{+0.041}_{-0.046}$ & $ 0.647^{+0.018}_{-0.018}$ & $ 0.659^{+0.015}_{-0.015}$ & $0.669^{+0.013}_{-0.013}$ \\
    $\tau$ & $0.087^{+0.016}_{-0.015}$ & $ 0.096^{+0.012}_{-0.016}$ & $ 0.098^{+0.018}_{-0.016}$ & $0.101^{+0.018}_{-0.017}$ \\
    $n$ & $0.965^{+0.022}_{-0.021}$ & $ 0.994^{+0.020}_{-0.017}$ & $ 0.996^{+0.020}_{-0.017}$ & $1.003^{+0.021}_{-0.018}$ \\
    ${\rm ln} [10^{10}A_s]$ & $3.083^{+0.037}_{-0.036}$ & $ 3.064^{+0.039}_{-0.037}$ & $ 3.064^{+0.040}_{-0.038}$ & $3.065^{+0.041}_{-0.039}$ \\
    $w_0$ & $<-0.629(68\% {\rm C.L.})$ & $<-0.943(68\% {\rm C.L.})$ & $<-0.978(68\% {\rm C.L.})$ &$<-0.977(68\% {\rm C.L.})$ \\
    $w_e$ & $-0.25^{+0.09}_{-0.05}$ & $ -0.33^{+0.03}_{-0.02}$ & $ -0.33^{+0.02}_{-0.02}$ & $-0.34^{+0.02}_{-0.02}$ \\
    $\xi$ & $<0.0026(68\% {\rm C.L.})$& $<0.0006(68\% {\rm C.L.})$ & $ <0.0005(68\% {\rm C.L.})$ & $<0.0005(68\% {\rm C.L.})$ \\
    \hline
\end{tabular}
\end{table*}

In order to get tighter constraint on $\Omega_m h^2$, we use the BAO
distance measurements \cite{BAO} which are obtained from analyzing
clusters of galaxies and test a different region in the sky as
compared to CMB. BAO measurements provide a robust constraint on the
distance ratio
\begin{equation}
d_z = r_s(z_d)/D_v(z)
\end{equation}
where $D_v(z)\equiv[(1+z)^2D_A^2z/H(z)]^{1/3}$is the effective distance  \cite{Eisenstein}, $D_A$ is the angular diameter distance, and $H(z)$ is
the Hubble parameter. $r_s(z_d)$ is the comoving sound horizon at the baryon drag epoch where the baryons decoupled from photons. We numerically
find $z_d$ using the condition $\int_{\tau_d}^{\tau_0}\dot{\tau}/R=1,R=\frac{3}{4}\frac{\rho_b}{\rho_{\gamma}}$ as defined in \cite{wayne}. The
$\chi^2_{BAO}$ is calculated as \cite{BAO},
\begin{equation}
\chi^2_{BAO}=(\vec{\textbf{d}}-\vec{\textbf{d}}^{obs})^T\textbf{C}^{-1}(\vec{\textbf{d}}-\vec{\textbf{d}}^{obs})
\end{equation}
where $\vec{\textbf{d}}=(d_{z=0.2},d_{z=0.35})^T$, $\vec{\textbf{d}}^{obs}=(0.1905,0.1097)^T$ and the inverse of covariance matrix read
\cite{BAO}
\begin{equation}
\textbf{C}^{-1}=\left(
                  \begin{array}{cc}
                    30124 & -17227\\
                    -17227 & 86977 \\
                  \end{array}
                \right).
\end{equation}
Furthermore, we add the BAO A parameter \cite{BAO_A},
\begin{eqnarray}
A&=&\frac{\sqrt{\Omega_m}}{E(0.35)^{1/3}}\left[\frac{1}{0.35}\int_0^{0.35}\frac{dz}{E(z)}\right]^{2/3}\nonumber \\
 &=&0.469(n_s/0.98)^{-0.35}\pm 0.017
\end{eqnarray}
where $E(z)=\frac{H(z)}{H_0}$ and $n_s$ are the scalar spectral index. In order to improve the constraints on the DE EoS $w$, we use the
compilation of  397 Constitution samples from supernovae survey \cite{SNeIa}. We compute
\begin{equation}
\chi^2_{SN}=\sum \frac{[\mu(z_i)-\mu_{obs}(z_i)]^2}{\sigma_i^2}\quad ,
\end{equation}
and marginalize the nuisance parameter.

We implement the joint likelihood analysis,
\begin{equation}
\chi^2=\chi^2_{WMAP}+\chi^2_{SN}+\chi^2_{BAO}.
\end{equation}
The cosmological parameters are well constrained. When the coupling between dark sectors is
proportional to the energy density
of DM, its constraint is much tighter than that from CMB data alone.  The likelihoods of the fitting results for the matter abundance,
parameters $w_0, w_e$ for DE EoS are also improved. Compared with the WMAP data alone, we see that the joint analysis by including
other observational data provides tighter constraints on the cosmological parameters.

\begin{figure}[hptb]
    \centering
    \includegraphics[scale=0.3]{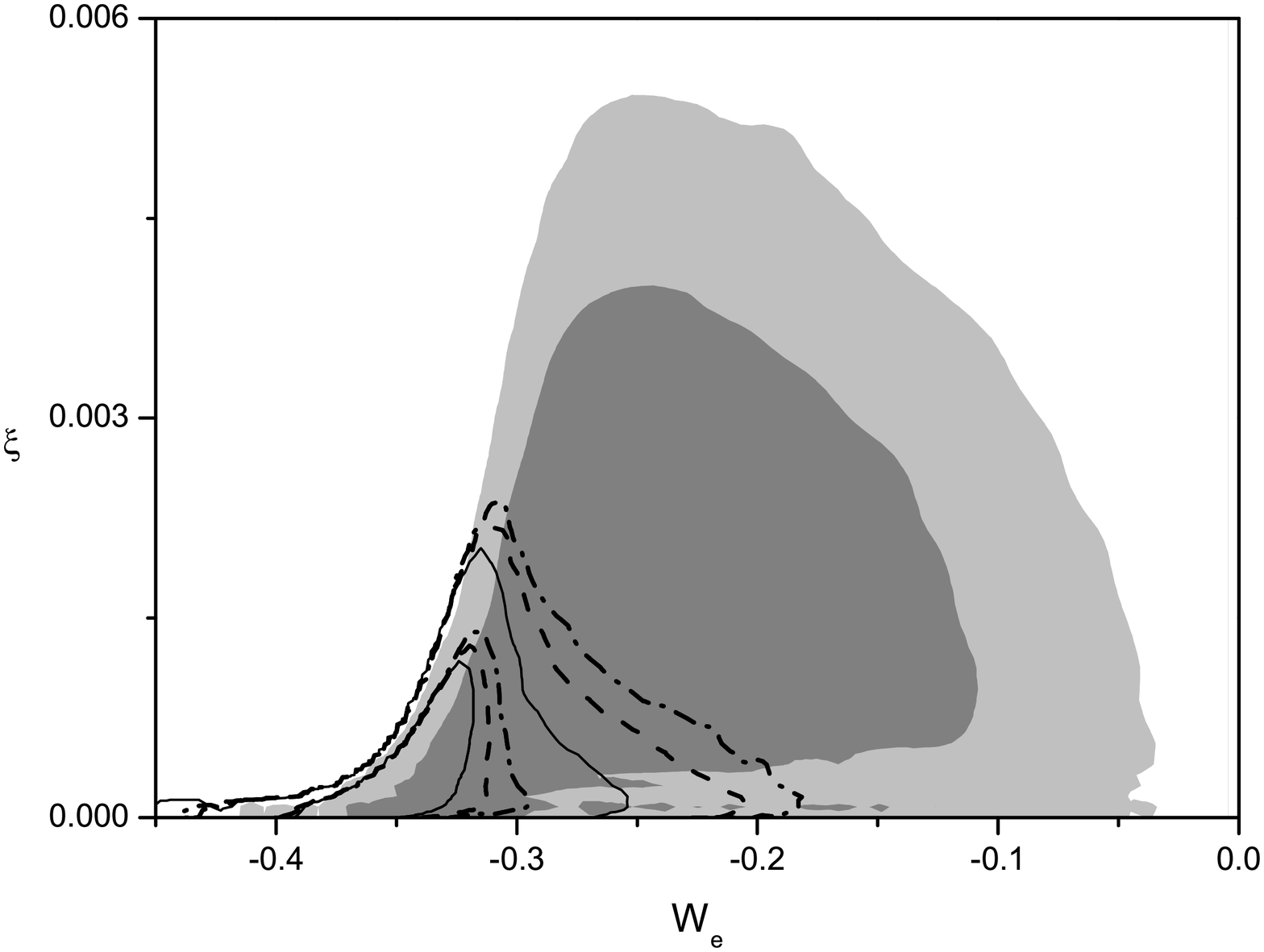}
    \caption{The 2-D marginalized confidence contours for $\xi$ and $w_e$. WMAP alone(in shades), WMAP+BAO(dash-dotted lines),
    WMAP+SNIa (dashed lines) and
    WMAP+BAO+SN data (solid lines).}
    \label{we.xi}
\end{figure}

In Fig.\ref{we.xi} we plot the 2d marginalized likelihood for the
interacting models from our MCMC run with CMB and other
observational data. The shaded regions are the result from WMAP
alone at $68\%$ and $95\%$ C.L. If we just look at the CMB result,
we see that when $w_e<-1/4$, the allowed $\xi$ becomes smaller when
$w_e$ decreases. When $w_e>-1/4$, the allowed $\xi$ is bigger. This
fitting result supports our theoretical observation obtained in the
stability analysis. The dashed and solid lines are the $68\%$ and
$95\%$ C. L. regions in combination of WMAP+SNIa and WMAP+BAO+SN
data sets, respectively. Fig.\ref{we.xi} summarizes our findings for
the possible relation between $w_e$ and the coupling constant from
data fitting.

\section{conclusions and discussions}

In this paper we generalized our previous study\cite{he2010} on the interacting DE model to the DE with time varying EoS. We reviewed the
perturbation theory and examined the stability in different model parameters. Based upon the perturbation formalism we have studied the signature
of the interaction between dark sectors from CMB angular power spectrum. Theoretically we found that there are possible ways to break the
degeneracies among the interaction, parameters in CPL parametrization of DE EoS and DM abundance. This can help to get tight constraint on the
interaction between DE and DM.

We have performed the global fitting by using the CMB power spectrum data from WMAP7Y results together with the latest SNIa, and BAO data to
constrain the interaction between DE and DM and other cosmological parameters. As anticipated from our theoretical analysis, the global fitting
can really break degeneracy of cosmological parameters and give tighter constraints on the interaction between dark sectors, DE EoS and DM
abundance.  The relation between $w_e$ and the coupling obtained from the fitting supports the result got in the stability analysis.

With the successful experience to deal with the interacting DE model with time varying DE EoS, our next step is to  extend our study to the field
theory based model to describe the interaction between dark sectors. The preliminary attempt was carried out in \cite{secondref}. More efforts
are required on this direction.

\begin{acknowledgments}
This work was partially supported by  NNSF of China and the National Basic Research Program of China under grant 2010CB833000.

\end{acknowledgments}

\end{document}